\documentclass[journal]{IEEEtran}
\ifCLASSINFOpdf
\else
\fi

\usepackage{lineno,hyperref}
\usepackage{booktabs}
\usepackage{color}
\usepackage{graphicx}
\modulolinenumbers[5]

\begin{document}
\title{CommuNety: A Deep Learning System for the Prediction of Cohesive Social Communities}

\author{Syed~Afaq~Ali~Shah,~\IEEEmembership{}
        Weifeng~Deng,
        Jianxin~Li,
        Muhammad~Aamir~Cheema
        and~Abdul~Bais~\IEEEmembership{}% <-this % stops a space
\thanks{S.A.A. Shah is with the Department of Information Technology, Mathematics and Statistics, Murdoch University, Australia,
e-mail: afaq.shah@murdoch.edu.au.}% <-this % stops a space
\thanks{W. Dong is with the University of Western Australia. J Li is with Deakin University, Australia. M.A. Cheema is with Monash University, Australia. A Bais is with University of Regina, Canada.}% <-this % stops a space
\thanks{Manuscript received April 19, 2005; revised August 26, 2015.}}

\markboth{}
{Shah \MakeLowercase{\textit{et al.}}}
% 

\maketitle

\begin{abstract}
Effective mining of social media, which consists of a large number of users is a challenging task. Traditional approaches rely on the analysis of text data related to users to accomplish this task. However, text data lacks significant information about the social users and their associated groups. In this paper, we propose CommuNety, a deep learning system for the prediction of cohesive social networks using images. The proposed deep learning model consists of hierarchical CNN architecture to learn descriptive features related to each cohesive network. The paper also proposes a novel Face Co-occurrence Frequency algorithm to quantify existence of people in images, and a novel photo ranking method to analyze the strength of relationship between different individuals in a predicted social network. We extensively evaluate the proposed technique on PIPA dataset and compare with state-of-the-art methods. Our experimental results demonstrate the superior performance of the proposed technique for the prediction of relationship between different individuals and the cohesiveness of communities. 
\end{abstract}


% \fi
\IEEEpeerreviewmaketitle

\section{Introduction}
\IEEEPARstart{W}{ith} the pervasiveness of low cost digital cameras and advent in computer vision and machine learning approaches, the collection and analysis of large image data has become a trivial task. As the value of photos is greatly determined by who appears in those photos (e.g., celebrity), labeling photos with their identities becomes an essential task \cite{Ortiz, sun2016social, weng2009rolenet, oro2017detecting}.

The popularity of social applications and social networking services (SNS) such as Facebook, Twitter, LinkedIn, Weibo, MOMO and Flickr has led to the formation of online social networks of users on these sites. At present, analyzing online comments (e.g., tweets) is a popular method to determine effective communities in social networks. Pfeil et al. \cite{Pfeil} proposed a study about the age differences of users in online social communities. They extracted information from MySpace’s user profile pages and divided the users into teenagers and older people communities. Users in the same community have common features, for example, teenagers have larger friends networks than older users. While text data contains rich information, it can be noted that the existing methods are unable to utilise the text data to get sufficient information about the social users. In addition, the social networks need to be more comprehensive and accurate \cite{zhao2017social}. With the advent of imaging technology and the availability of portable high resolution cameras such as on smartphones, users can now upload their images and profiles to social media websites and share photos with other users who are part of their social community \cite{xu2018trust, cheung2019detecting}. Social media users upload countless photos of social activities each day, and the relationship among those who appear in these photos cannot be mined accurately only from text data. Hence, defining online social networks with user-uploaded images, and extraction of human features, such as faces or body from photos becomes an important procedure in building social networks \cite{garg2019hybrid, lu2016tag, zhang2019personalized}. Note that the popular SNS applications have very large user bases. In 2018 alone, Facebook had 2.2 billion monthly active users. Flickr had over 90 million monthly users, and the number of monthly users of Weibo exceeded 0.44 billion \cite{Dmr}. Therefore, the mining of a potential relationship between social network users is a challenging problem.

To overcome the above challenges, in this paper, we propose a deep learning system, CommuNety, which uses image data for the prediction of comprehensive and cohesive online social communities. The proposed system complements existing works and is helpful in discovering communities when there are no explicit relationships (e.g., discovering communities in an image database) or discovering communities when not all relationships are directly represented in the network e.g., two people may not be friends on social media and may have never interacted on the platform, however, if they appear together in some photos, they have a relationship which can only be discovered using images.

Several deep learning algorithms have been developed in recent years and have achieved significant breakthroughs in image recognition tasks. In 2014, Simonyan and Zisserman proposed a Deep Convolutional Neural Network (CNN) architecture and achieved an outstanding classification performance \cite{Simonyan}. Parkhi et al. applied the VGG (Visual Geometry Group) network structure to face recognition task and achieved results comparable to other face recognition techniques  \cite{Parkhi}. Razavian et al. have demonstrated that the features extracted from CNN are powerful and the models trained using CNN features have superior performance. Such features can be used for visual recognition tasks \cite{Raza}. 

Inspired by prior approaches, in this paper, we propose a high-performance face recognition model, which learns distinctive image features. The proposed model is then used to predict community network and its hierarchy that is centered at the target person. In our proposed technique, every photo in the training set is also ranked using the term frequency inverse document frequency (TF-IDF) numerical statistics. Then, the strength of relationship between each pair of persons in the photos is represented by the sum of the TF-IDF values of their group photos. As a result, the social network outputted from the proposed prediction system contains all the persons who have direct or indirect relationship with the target person and different relationship strength among them. 

Recognizing people from high-quality photos, which contain high-resolution facial images, is a trivial task for humans. However, well trained autonomous system still struggle with this challenging task. This is because of the variations in natural images, such as changes in illumination and viewpoint change or head rotation. Moreover, although some progresses have been made recently in recognition from a frontal face without face location, non-frontal views are more common in social media photo albums. A few face recognition techniques perform face detection as a preliminary step \cite{shah2017efficient, hu2019disentangled, zhang2018dynamic}. Note that face detection can be regarded as a two-class (face versus non-face) classification problem. However, these techniques cannot deal with significant variations in face images such as head rotation and view changes, etc. to detect and recognise faces. Other model-based approaches require that the initial locations of faces are known in advance \cite{Hsu} and then they perform face tracking to recognise individuals in the image data. This paper overcomes these challenges. The significance of this research is to recognise people from any viewpoint and associate them with established cohesive social communities \cite{Zhang}. 

The contributions of this paper can be summarised as follows: 
\begin{itemize}

   \item \textbf{First}, we propose a deep learning model to predict cohesive social communities or networks using image data and face recognition.
   \item \textbf{Second}, we propose a novel algorithm to calculate the relationship strength among people in the predicted social network. The improved social networks are quite informative. We also present the final social networks using data visualization techniques. 
      \item \textbf{Third}, we propose novel features for image-generated networks compared with other social communities formed in social media.
   \item \textbf{Four}, we perform extensive evaluation of the proposed technique. Our experimental results demonstrate the superior performance of the proposed system on the PIPA (“People In Photo Albums”) dataset. 
\end{itemize}

The rest of this paper is organised as follows. The next section discusses the prior works related to this research. Section \ref{sec:PM} describes the proposed methodology and provides information about the training of face recognition model, construction of social networks, and analysis of the predicted social networks. The PIPA dataset used for the evaluation and data pre-processing are discussed in Section IV. Section V presents our experimental results. The paper is concluded in Section VI.

\section{Literature Review}
Chen et al.\cite{Chen} proposed a technique to identify family and non-family images, which were collected from social media and to predict the pairwise relationship of persons who were in the same family images. To categorize different group types or events, a bag-of-face-subgraphs (BoFG) was proposed. BoFG contained meaningful subgraphs, which represented a group photo, and the occurring frequency of these subgraphs was adequate to identify specific image types. The authors trained an SVM classifier using BoFG features and their technique achieved an accuracy of 89$\%$ on family image recognition. In addition, a Naive Bayesian classifier was used to predict the pairwise relationship by getting the image frequency of appearance of the informative subgraph in the image collections. Their proposed technique achieved good improvement over prior works, especially in image categorization area. However, there are still several limitations of their technique. For instance, the images used in the training and testing phases are frontal face images. Hence, if BoFG is applied to open world images, which contain large number of non-frontal facse, the performance would be significantly affected. Moreover, in their proposed method, the pairwise relationship is identified based on the gap of age and gender in a household. This special feature is not feasible for other types of relationship, which do not involve age and gender gap. This limits the application of their proposed technique on real world social network data.

Kim et al. \cite{Kim} developed an associative network structure called Face Co-occurrence Networks (FCON), which was used to recommend reliable social friends and explore relationships among people based on tagged personal photos. FCON consists of vertices (V) and edges (E), where V is a set of faces, which appeared in photos and E is a set of links between each pair of faces (aka. co-occurrences of faces), both V and E are accumulated. Converting all photos into a global FCON, the weights of V and E in the network were obtained by accumulating V and E in each subnetwork. Subsequently, parts of weights which were related to the target user were calculated to get a set of scores and compare these scores with a pre-set threshold. Finally, using the vertices which have scores higher than the threshold to establish a target user-centric relationship network. Besides, the authors also develop a web-based system named VizFaceCo for data visualization.  An aspect that is obviously worth improving is that their technique does not include face recognition. The photos are manually annotated with corresponding names before building FCON. In contrast, in our proposed technique, automatic face recognition is used as a core technology.

Oh et al. \cite{Oh} proposed an optimized model for person recognition, called naeil2, which is capable of handling large variations in person images. naeil2 consists of seventeen cues (including five vanilla regional cues, two head cues, ten attribute cues) and DeepID2+ face recognition module. All the cues were obtained from the seventh layer (fc7) of AlexNet \cite{Oh}, and concatenated together. Finally, these cues and DeepID2+ using L2 normalization was combined to build the final naeil2 model. 

naeil2 has been shown to achieve an outstanding person recognition performance. However, this model relies on multiple features such as several body cues to identify persons. In most of the social media photos, multiple body cues are hard to capture and therefore their proposed technique fails in these situations.
% \begin{figure*}[!htb]
% 	\includegraphics[scale=0.45]{vgg-face-architecture.png}
% 	\caption{Architecture of the proposed CommuNety model for the prediction of social communities.}
% 	\label{fig1}
% \end{figure*}

Dong et al. \cite{Dong} proposed a method for human age identification. They proposed CNN based DeepID architecture. In their method, the loss function for classification was modified and a distance term was added to the loss function to emphasize on the relationships between labels. They used different parts of face images to train multiple classifiers, and by comparing the accuracy of an exact match (AEM), the eye region was found to be the most significant feature, which can reflect the age of the person. To further improve AEM, different models were combined, and the best model combination was shown to achieve good performance.

They also described in detail the transfer learning strategy adopted in their work that used fewer data samples to train their model to achieve good performance. Concretely, they used large-scale data sets to train a face recognition model, then transferred the parameters of convolutional layers to another network which had same architecture, but the parameters in fully connected layers were randomly initialized. This new network was fine-tuned using the small-scale dataset to get the desired age classification model.

The limitation of their technique is that the performances of the face recognition models were not outstanding compared to the state-of-the-art face recognition techniques e.g., the lowest error rate for DeepID is 0.4. In other words, the accuracy of the best model is 60$\%$ \cite{Dong}. One of the reason is that the architecture of CNN used in DeepID is relatively simple, for instance, the DeepID only has four convolutional layers. Their model was not able to handle the image data complexity, therefore it was under-fitting. The accuracy of recognition can be improved by using a more complex CNN architecture and more training data \cite{Geron}. 
%Parkhi et al., \cite{Parkhi} proposed a deep face recognition architecture called VGG-Face to extract discriminating face features for face recognition task. The architecture of VGG-Face network is shown in Figure~\ref{fig1}. It is composed of sixteen blocks, the first eleven blocks are convolutional layers, the last three blocks are called Fully Connected (FC) layers. Each block followed by one non-linear activation function ReLU, and five max-pooling layers are interspersed between blocks to reduce computational load. Input size of this network is 224x224 pixels. Their proposed architecture achieved an outstanding prediction accuracy on face recognition. 2.6 million face photos were used for training the VGG-Face, such a huge amount of training data provides fairly appropriate parameter setting for this model. Their proposed technique can also fit small data tasks through the transfer learning strategy. 
In this paper, we overcome the limitations of the prior methods and propose a novel technique for the prediction of social communities using images and calculate the relationship strength between the social media users.

\section{Proposed Methodology} \label{sec:PM}
In this section, we describe our proposed deep learning based system, CommuNety, to predict social community centered on the input image. We propose a novel algorithm to calculate the relationship strength between people in the predicted social community. The proposed deep learning pipeline consists of two phases including the face recognition phase and, the community prediction and formation phase. In face recognition phase, our proposed deep learning model is trained to perform accurate face recognition. In the community prediction and formation phase, we develop a novel Face Co-occurrence Frequency algorithm and calculate the relationship strength to predict communities. In addition, we also propose novel features for the predicted communities by analyzing their properties.

\subsection{Face Recognition Phase}
To establish accurate and cohesive social networks, the most challenging task is to accurately recognize persons in given photos. We first detect faces in the input images by using the Viola and Jones algorithm \cite{viola2004robust}. The outcome of face detection (i.e., the bounding box of faces) is validated by the annotations provided in \cite{Zhang}. The detected face images along with their labels are then fed to our deep learning based face recognition system. These images are used for the training and testing of our proposed deep neural network, which is discussed in the following section.

\subsubsection{Deep Neural Network Architecture}
We propose a deep face recognition architecture to extract discriminating features for face recognition task. The proposed deep learning architecture is composed of sixteen blocks. The first eleven blocks consist of convolutional layers. Each block followed by one non-linear activation function ReLU, and five max-pooling layers are interspersed between blocks to reduce computational load. The last three blocks are called Fully Connected (FC) layers. The last layer is a softmax layer for multi-class classification and its dimension is equal to the number of class labels in task. %Table \ref{table1} provides a summary of the most representative layers of the proposed deep learning architecture.

\subsubsection{Deep Network Training and Testing}
The neural network is trained as a multi-class classifier to recognize persons using their face images. The class probability is computed using the following equation, which computes probability in the range between 0 and 1 for each class:

\begin{equation} 
y_{j} = \frac{ e^{ x_{j} } }{ \sum_{i=1}^{N}e^{ x_{i} } } 
\label{eq1}
\end{equation}
where \emph{N} represents the number of classes and $y_{j}$ is the probability of class \emph{j}. $x_{j}$ is the output of the \emph{jth} neuron in the soft-max layer. Its role is to increase the probability of true class label. In addition, we use cross-entropy loss function as in Eq.~(\ref{eq2}) for the softmax layer: 
\begin{equation} 
L = -{\sum_{j=1}^{N}1[al=j]\log y_{j}}
\label{eq2}
\end{equation}
where \emph{al} is the actual label of input.

During testing, given a test face image, the network then predicts the class label for the input test image. The output of face recognition is then used in the subsequent modules and to predict the social community as discussed below.

\subsection{Community Prediction and Formation Phase}
Once the faces have been successfully recognised in the input images, the next phase is to predict the social communities using facial images. We propose two algorithms to predict social communities using our face recognition system, and compute relationship strength for each pair of connected nodes in the communities.

\subsubsection{Recursive Face Co-occurrence Frequency}
Our proposed algorithm to predict communities is similar to FCON \cite{Kim}, however, it is recursive in nature. A dictionary is first defined to store face co-occurrence frequencies as follows:
\begin{equation} 
F_{i} = \{P_{1}: f_{1}, P_{2}: f_{2}, ..., P_{k}: f_{k}\}
\label{eq3}
\end{equation}
where \emph{i} is the target candidate, key $P_{k}$ is the class of the \emph{kth} person in the dataset, and key value $f_{k}$ represents the number of times the \emph{kth} person appears in the given album. The key values are initially zero.
\begin{figure*}[!htb]
\begin{center}
	\includegraphics[scale=0.6]{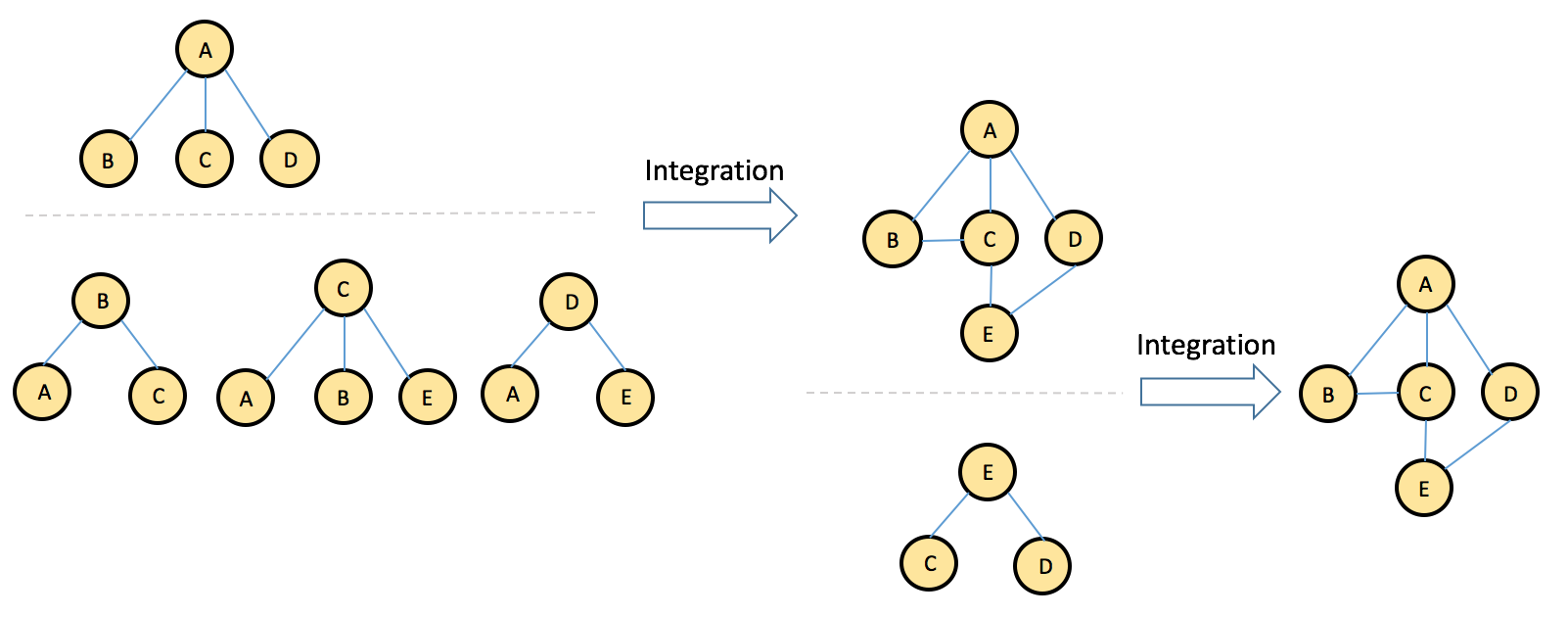}
\end{center}
\caption{Social Network Construction. \textbf{Left:} Person A is the root, which is directly connected to persons B, C and D. \textbf{Right:} The final network has two layers, as only person E is on the latest layer and E does not connect to any new person who is not in the current network.}
\label{fig2}
\end{figure*}

Given an input target person's face image, our proposed face recognition system recognizes and collects all the photos of the target class. Next, comparison of the photo labels of other classes with the collected target photo labels is performed. The images of other classes are input to face recognition system for predicting the class label of the input image. When the input person's class is predicted, the key value corresponding to the person's name in the co-occurrence frequency dictionary is incremented by one. After assigning all the matched photos to the dictionary, persons whose key values are larger than zero are considered as directly connected with the target in the social network. Below, we provide a definition of elements contained in the predicted social network. 

\textbf{Definition}. Root, nodes, and layers:
\begin{enumerate}
    \item Initial target is the root of its social network, meanwhile, it is on layer 0.
    \item Other people in the social network are nodes. Nodes that are directly connected with the root are on layer 1; similarly, the nodes on the second layer are connected to the nodes on the first layer.
    \item Because multiple persons may occur in a group photo, therefore the nodes on the same layer may be connected.
\end{enumerate}

Each person on layer 1 is treated as a new target and the same method (as stated above) is followed to build their corresponding single-layer social network. The new community network is then integrated with the previous network to build a 2-layer social network. We only add people who do not already exist in the previous network to the second layer. This process is repeated until no more new person can be found to join the network, and finally a complete social network centered on the initial target person is set up. Figure~\ref{fig2} shows an example of social network construction process. Person A is the root, which is directly connected to persons B, C and D. The final network has two layers, as only person E is on the latest layer and E does not connect to any new person who is not in the current network.

\subsubsection{Prediction of Relationship Strength in Social Communities}
To predict the relationship strength of persons in social communities, we propose an image ranking algorithm to assign scores to the images that determine the relationship strength in a community. To achieve this, TF-IDF is used for ranking in the predicted community. 

TF-IDF is a statistical analysis technique for weighing that reflects the importance of a word for the documents in a corpus. This importance is obtained by comparing the relative frequency of a word in a particular document with the inverse proportion of the word in the entire corpus \cite{Ramos}. In the proposed technique, the corpus consists of all the group photos, where each photo represents a document in the corpus, and the words are replaced by persons in the group photos.

In the proposed technique, the formula for TF-IDF is defined as follows. Given the group photo set \emph{G}, a candidate \emph{c}, and a single group photo \emph{g} $\in$ \emph{G}, the TF-IDF is represented as: 

\begin{equation} 
TF-IDF_{c,g} = f_{c,g} * \log (|G|/f_{c,G}) 
\label{eq6}
\end{equation}
where $f_{c,g}$ is the number of times \emph{c} appears in \emph{g}, $|$G$|$ represents the size of the group photo set, and $f_{c,G}$ equals the number of group photos in which \emph{c} appears in \emph{G}. 

The TF-IDF formula can be separated into two terms TF and IDF as follows:

\begin{equation} 
TF_{c,g} = f_{c,g}
\label{eq7}
\end{equation}

\begin{equation} 
IDF_{c,G} = \log (|G|/f_{c,G}) 
\label{eq8}
\end{equation}

Since a given person in each photo can only appear once, therefore $TF_{c,g} = 1$. Meanwhile, each candidate has their own fixed IDF value, as the number of times they appear in the entire photo collection is fixed. The more a person appears in the photo collection, the smaller the IDF they receive, and is considered as the lesser important in a specific photo.

The group photos are ranked by calculating the averages of the TF-IDFs of all candidates in each photo:  
\begin{equation} 
Score_{g} = \frac{\sum_{i=1}^{k} TF-IDF_{i,g}}{k} 
\label{eq9}
\end{equation}
where \emph{k} represents the number of persons in \emph{g}.

Intuitively, the score of a photo is related to the IDF values of the persons in the photo. For example, if the persons in a photo appear only a few times in the entire collection, the score of this photo is high. On the contrary, if most people in a photo appear in the photo collection many times, then the IDFs of these people are small, and the significance of this photo is low.

Ultimately, the strength of the relationship between each pair of connected people in the social network is represented by the sum of the scores of all of their photos in which both persons appear. Each edge in the social network is assigned a weight representing the strength of the relationship (larger the better) between the two persons connected by the edge. %Larger numbers on the edges represent closer relationships of the corresponding node pairs and vice versa.

\section{Image Data for Evaluation}
\subsection{Dataset:}
The proposed technique is evaluated on People In Photo Albums (PIPA) dataset. PIPA dataset contains 37107 Flickr personal photo album images, with 63188 head annotations of 2356 identities, all the images have Creative Commons Attribution License \cite{Zhang}. The dataset is divided into train, val, test, and leftover sets, with a proportion of 45$\%$, 15$\%$, 20$\%$, 20$\%$, respectively. We used the same experimental protocol as in \cite{Oh}, hence the same image sets are used in this paper.%, which contains 30552 photo images, with 51751 head annotations of 2356 identities. 

The train, val and test sets in the dataset contain distinct identities i.e., the class labels in training and test set were totally different. Therefore these image data cannot be used directly for the proposed technique. Besides, the number of photos from different identities varies significantly, e.g., the minimum number was only 5, and such a small data size is not enough to train the proposed model. Therefore, we pre-process the data for our deep learning model.

\subsection{Data Pre-processing}
In data pre-processing stage, data cleaning, redistribution, and data augmentation are performed.

\begin{figure}[t]
\begin{center}
    \includegraphics[scale=0.15]{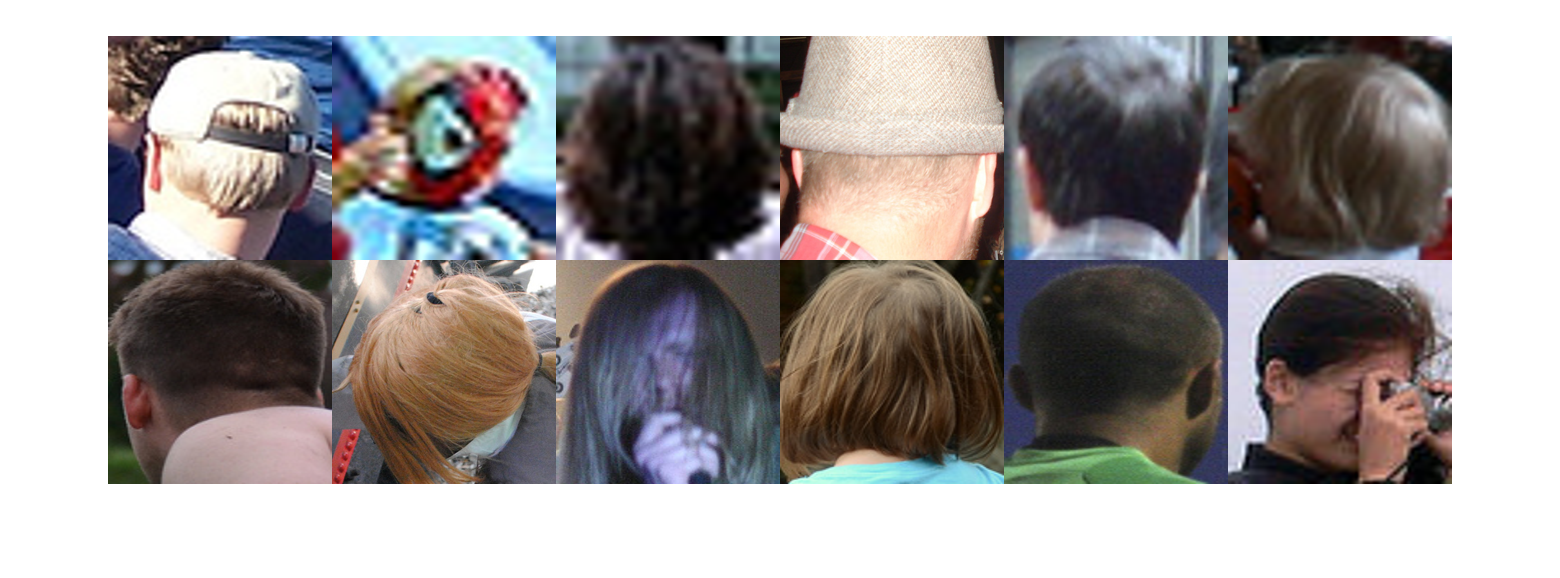}
	\end{center}
	\vspace{-2mm}
	\caption{An Example of Poor Quality Instances i.e., Head Images.}
	\label{fig3}
\end{figure}

\begin{figure}[!htb]
	\includegraphics[scale=0.35]{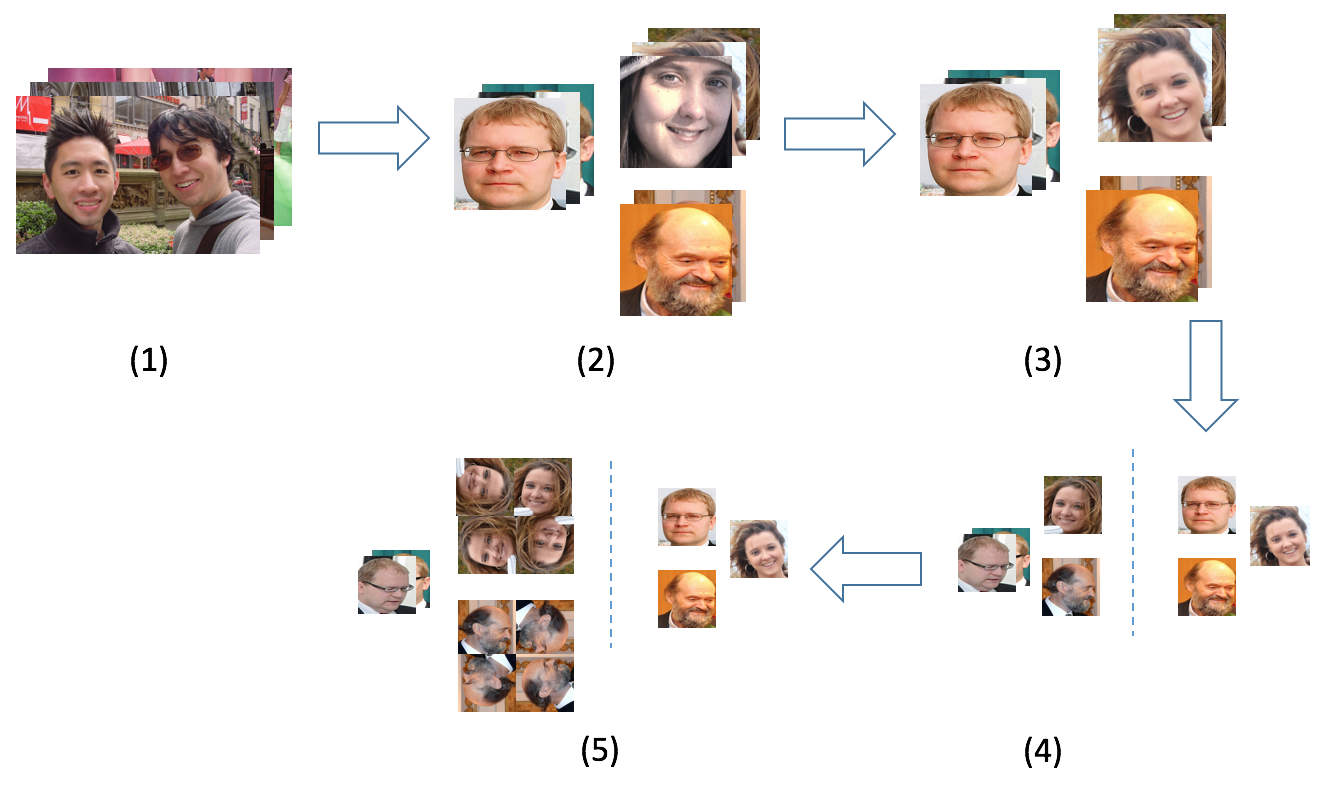}
	\caption{Data Pre-processing. \textbf{(1)} Personal photos in PIPA Dataset. \textbf{(2)} Re-sized head images cropped from original photos. \textbf{(3)} Good quality head images. \textbf{(4)} Training and test data. \textbf{(5)} Augmented training data. 
	}
	\label{fig4}
\end{figure}

First, we crop all the face images of identities in the train, val and test sets. All instances are then resized to 224x224 to fit the input size of the proposed model. Second, data cleaning is performed. The cropped images have different appearances, including the front face, the side face, and even the back of heads. The back of head does not contribute in recognition and could affect the training of our deep network, therefore, these images are removed from the dataset. An example of these images/instances is shown in Fig~\ref{fig3}. Third, the training and test images are randomly selected with the proportion 80$\%$ and 20$\%$, respectively. Because the instance size of each class is different, the stratified sampling method is used for data allocation to avoid significantly biased results \cite{Geron}. The last step is to perform data augmentation. We set 8 as the minimum number of instances per class. For the classes with insufficient instances, we perform data augmentation by rotating their instances by different angles, flipping and scaling. As a result of this, the numbers of instances in those classes are expanded. Figure~\ref{fig4} presents different steps involved in our data pre-processing.

After pre-processing and data augmentation, the dataset has 2356 classes, training set contains 41533 face images out of which 8613 of them are augmented. The test set contains 8230 face images. The distribution of images in the pre-processed dataset is shown in Table~\ref{table3}.

\begin{table}
\centering

\begin{tabular}{@{}cccc@{}}
\toprule
Split& All& Train& Test \\ \midrule
Instances (augmentation)& 49763& 41533 (8613)& 8230  \\
Identities& 2356& 2356& 2356\\
Average identity& 21.12& 17.63& 3.49 \\ \bottomrule

\end{tabular}
\caption{Statistics of the pre-processed dataset}
\label{table3}
\end{table}

\section{Experimental Results}
In the following, we first train the proposed model for the face classification task, then construct the desired social networks. 

%\subsection{Face recognition task}
During training, we use Stochastic Gradient Descent (SGD) and back propagation to decrease the Loss Function. SGD randomly chooses one training instance at each step and calculates the gradients based only on that single instance. This speeds up the algorithm as it only manipulates little data at each iteration, especially on huge training sets \cite{Geron}. In addition, to find the most satisfactory gradient, the learning rate is set to gradually decrease in the range of 0.005 to 0.00001 as the number of epochs increases. The learning rate changed every 30 epochs on average.

%\subsubsection{Performance of Face Classification Task:}
The model is trained to solve the multi-class classification task. It is assessed by top 1 error rate of classification. We compared the highest probability class of each sample with the actual classes, the top-1 error reflects the proportion of the number of incorrectly predicted samples to the total number of input samples. 

%The error rates of the classifier in training process are shown in Figure~\ref{fig5}, the vertical axis represents the model's error rates on test set in the corresponding epoch.  
\subsubsection{\textbf{Comparison with State-of-the-art}}
We compare our proposed technique with the state-of-the-art methods including the deep learning model naeil2 \cite{Oh} and DeepFace \cite{brunelli1995person}. Our experimental results are reported in Table \ref{table4a}.

\begin{table}
\centering
\begin{tabular}{c|c}
\toprule
Method& Accuracy \\ \midrule
naeil2 \cite{Oh} & 83.88 \\
DeepFace \cite{brunelli1995person} & 46.66 \\
\textbf{Proposed Technique} & 86.87 \\
 \bottomrule

\end{tabular}
\vspace{2mm}
\caption{Comparison of the proposed technique with the state-of-the-art methods.}
\label{table4a}
\end{table}

As can be noted, the proposed model's classification accuracy is 86.87$\%$, i.e., the top-1 error rate is 1-0.8687 = 0.1313. naeil2 \cite{Oh} fine-tuned the pre-trained AlexNet model using head images in PIPA Dataset and achieved accuracy of 83.88$\%$ \cite{Oh}. DeepFace achieved an accuracy of 46.66\% on PIPA dataset. These results demonstrate the superior performance of the proposed technique, which relies on face images to predict social communities.

% \begin{figure}[!htb]
% \centering
% 	\includegraphics[scale=0.25]{380epochs.png}
% 	\caption{Learning process of face classification task.}
% 	\label{fig5}
% \end{figure}

\subsubsection{\textbf{Implementation details}}
Our technique is developed in MATLAB. All our experiments have been performed on a machine with Intel Corei5 CPU and 16GB RAM. % Due to the transfer learning strategy, all the features of images are computed using the first fully connected layer (fc6) of VGG-Face pre-trained model \cite{Parkhi}. The input image size is 224 x 224 x 3. After feature extraction, all the image data are replaced by the corresponding CNN features with a size of 1 x 1 x 4096. We construct the three layers Neural Network to fit the training features.

\subsection{Community Prediction and Formation:}
\begin{figure}[t]
\centering
	\includegraphics[scale=0.38]{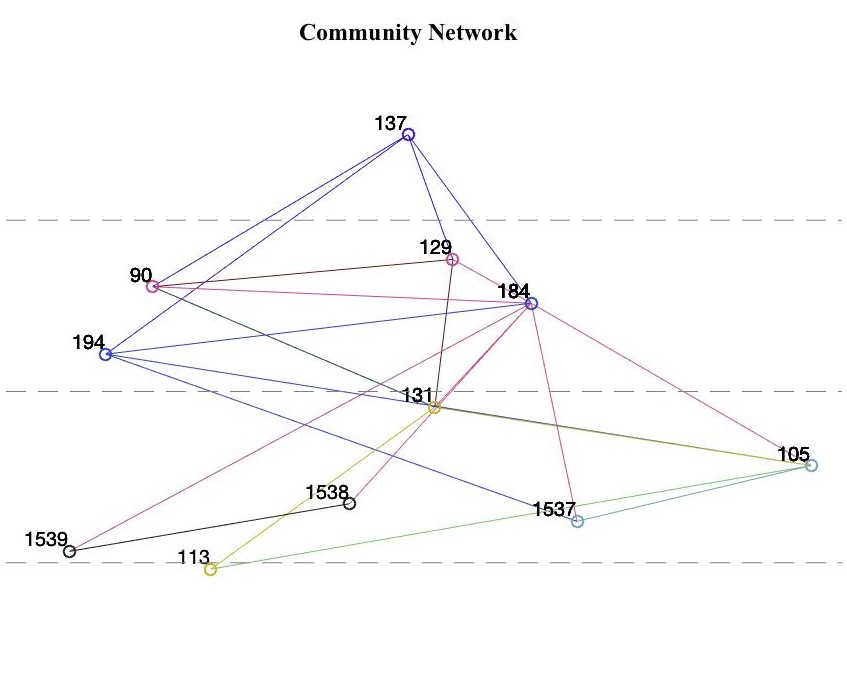}
	\vspace{-3mm}
	\caption{An Example of a Predicted Social Network.}
	\label{fig6}
\end{figure}

For social networks construction task, a complete social network prediction system is devised that is built on top of our face recognition model. The proposed system is evaluated on PIPA dataset. The input to social network prediction system is a face image of the target person, and a predicted social network graph starts with the target person as its node as shown in Figure~\ref{fig6}. Figure~\ref{fig6} is divided into three parts by the dotted line, where each part represents a layer. The numbers beside the nodes are people's identity, e.g., person 137 is the target, and they are also the root of this network. Moreover, to enhance visibility, the edges emitted from the same node have the same color. 
%The proposed model is responsible for recognizing the input image and all the other candidates' photos that should be included in the social network. As the input data of the model is CNN features of images, so all images related to the target need to be converted into CNN features with the size of 1x1x4096 using the method described in Section 4 before recognition. 

\begin{figure}[t]
\centering
	\includegraphics[scale=0.22]{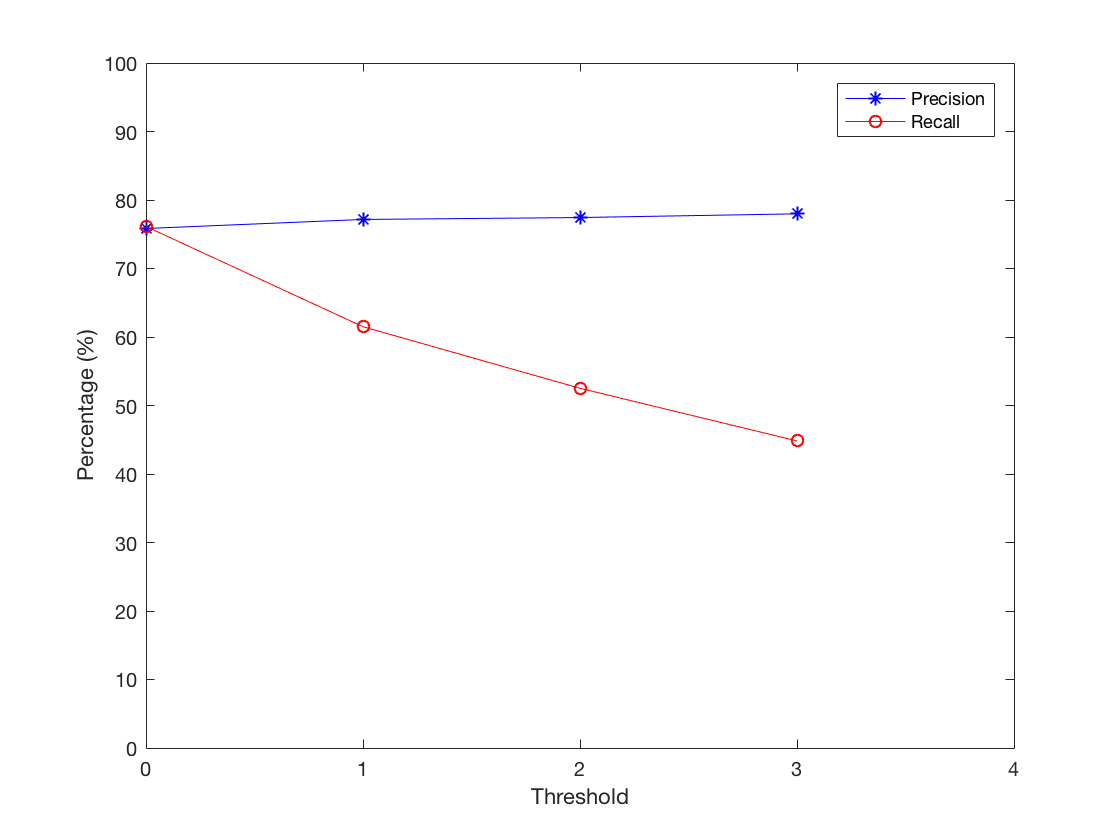}
	\caption{Precision and Recall versus the Minimum Frequency Threshold.}
	\label{fig7}
\end{figure}
\begin{figure*}[t]
\centering
	\includegraphics[scale=0.52]{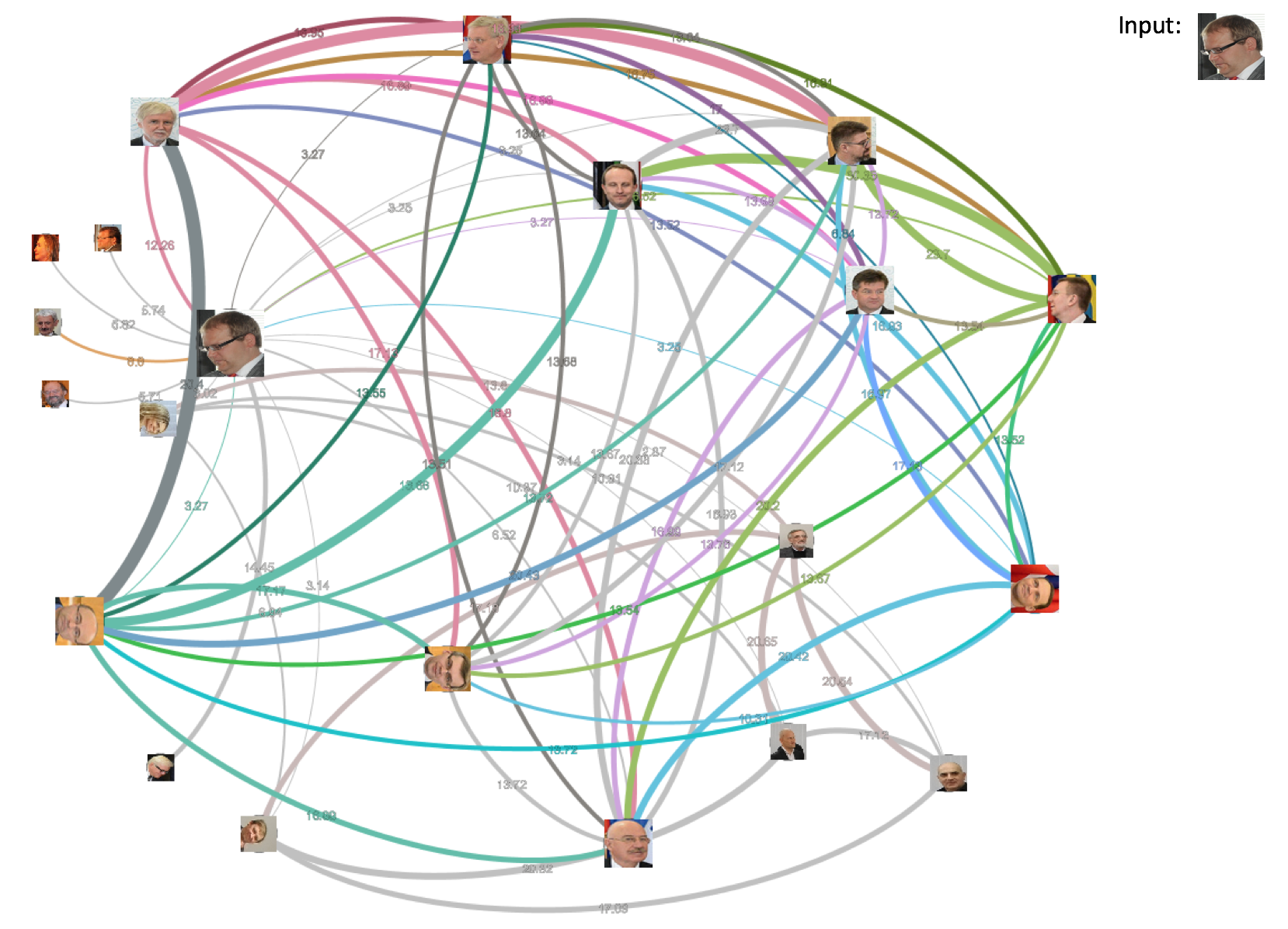}
	\caption{Social Network Centered at Person ID 1.}
	\label{fig8}
\end{figure*}
\begin{figure*}[!htb]
\centering
	\includegraphics[scale=0.52]{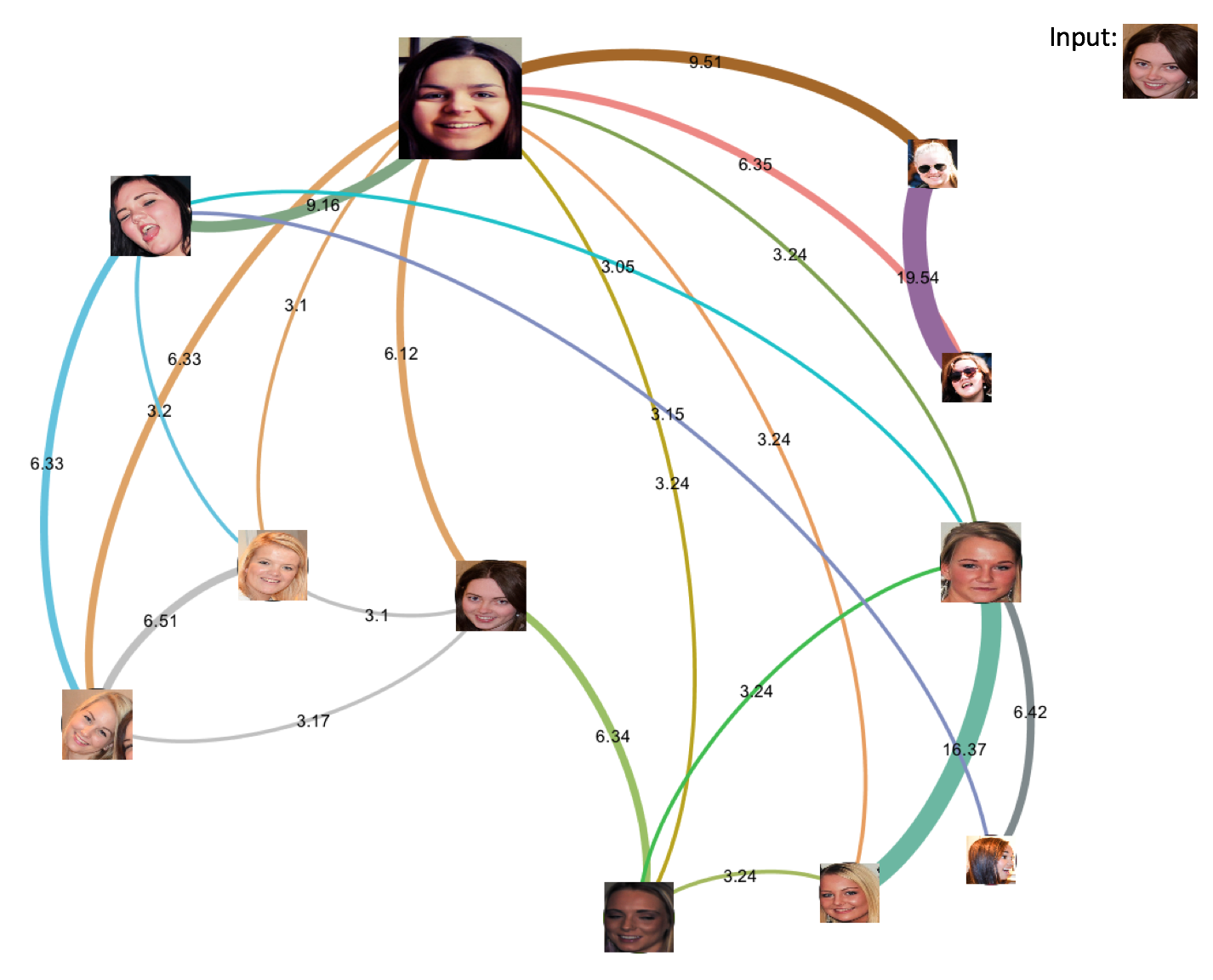}
	\caption{Social Network Centered at Person ID 137.}
	\label{fig9}
\end{figure*}

\subsubsection{Performance of Community Formation Task}
Precision and Recall criterion is used for evaluating all predicted networks. For each predicted label (predicted person name) in a network, it can only be judged whether it is consistent with the true label, no matter which class it belongs to. Thus, the multi-class classification tasks are converted into binary ones. Every predicted label is recorded as one of true negatives (TN), false positives (FP), false negatives (FN), or true positives (TP) base on the classification result. 

Precision is the accuracy of the positive predictions \cite{Geron}, the equation is shown as:

\begin{equation} 
Precision = \frac{TP}{TP+FP}
\label{eq4}
\end{equation}
where TP represents the number of people who exist in the networks and are correctly predicted. On the contrary, FP are wrongly classified person into the networks. However, precision is deceptive in some cases, for example, predicting one person as TP and to ensure that is correct, the precision is equal to 100$\%$, but the network could not be constructed by the single person. Hence, precision is necessarily utilized along with recall, a.k.a. true positive rate (TPR). As Eq.~(\ref{eq5}) shows, FN represents the number of persons who should be in the social networks and are not there.
\begin{equation} 
Recall = \frac{TP}{TP+FN}
\label{eq5}
\end{equation}

The precision may be improved by setting a minimum face frequency. However, this negatively affects recall. In Figure 5, we study this trade-off and observe that precision does not improve much whereas recall is severely  affected. Thus, we set minimum face frequency to be zero for the rest of the experiments. Only those individuals whose face frequency is greater than this threshold are classified into the corresponding social network. 

%Evaluation is still implemented base on the pre-processed training set and test set. Because the head images have been assigned to the corresponding candidates' folder, and the labels have been set, so we can find people who actually appear in the same photo, which are regarded as actual value to compare with predicted one layer networks. To find out the optimal Precision and Recall combination, the minimum face frequency is variable. 
Figure~\ref{fig7} shows the precision and recall for different thresholds and frequencies. When the threshold increases, the precision is not significantly improved, however, the recall is greatly reduced. We therefore empirically set the threshold to 0. The achieved precision with this threshold is 75.84$\%$, and the recall is 76.13$\%$. %In the following study, the threshold is fixed to 0.

\subsection{Prediction of relationship strength between candidates and analysis of social network properties}
To explore the relationship strength between candidates, we first calculate the IDF of each person using Eq.~(\ref{eq8}). The score of a photo is represented by the average of IDFs of all candidates in the photo. Once all the IDFs have been computed, the number of persons in each photo is calculated. The statistics of IDF and photo score is shown in Table~\ref{table4}.
\begin{table}
\centering
\begin{tabular}{@{}cccc@{}}
\toprule
Type& Min& Max& Average \\ \midrule
IDF& 2.45& 4.35& 3.31  \\
Photo Score& 2.45& 4.35& 3.08\\ \bottomrule
\end{tabular}
\vspace{2mm}
\caption{statistics of IDF and photo score}
\label{table4}
\end{table}

As discussed in Section 3.2, the relationship strength between two candidates is obtained by summing the scores of photos in which both appear together. Hence, we improve our social network prediction system by enabling it to record the face co-occurrence frequencies and corresponding photo names simultaneously. Then, find the scores of those photos from the previously defined photo score library to calculate the relationship strength.

All scores that reflect the relationship strengths are then displayed in the final social network graph. The example plots are shown in Figure~\ref{fig8} and Figure~\ref{fig9}. These social networks are built with two different target persons, respectively. The nodes are replaced by candidates' face images for visualisation, moreover, in addition to the scores on the edges, the thickness of the edges also reflects different relationship strength. 

\subsection{Analysis of Predicted Social Community}
In this section, we analyze the whole social community set after inputting all the test data to our proposed system and keeping all distinct communities. By counting the size distribution of social communities and the density of the communities, the cohesiveness of these predicted communities using image data is achieved. Community size and density are two of the primary network properties. Community size represents the number of nodes in a community. Community density refers to the Actual Connection-Maximum Connection ratio of a community as in Eq.~(\ref{eq10}):
\begin{equation} 
D(s) = \frac{m_{s}}{n_{s}(n_{s}-1)/2} 
\label{eq10}
\end{equation}
where $m_{s}$ is the number of edges in social network $s$, $n_{s}$ represents the amount of nodes in the network. The more the edges, the denser is the community. A community is more cohesive when it has larger density and smaller size \cite{brunelli1995person, Candan}. 

\begin{figure}[t]
\centering
	\includegraphics[scale=0.25]{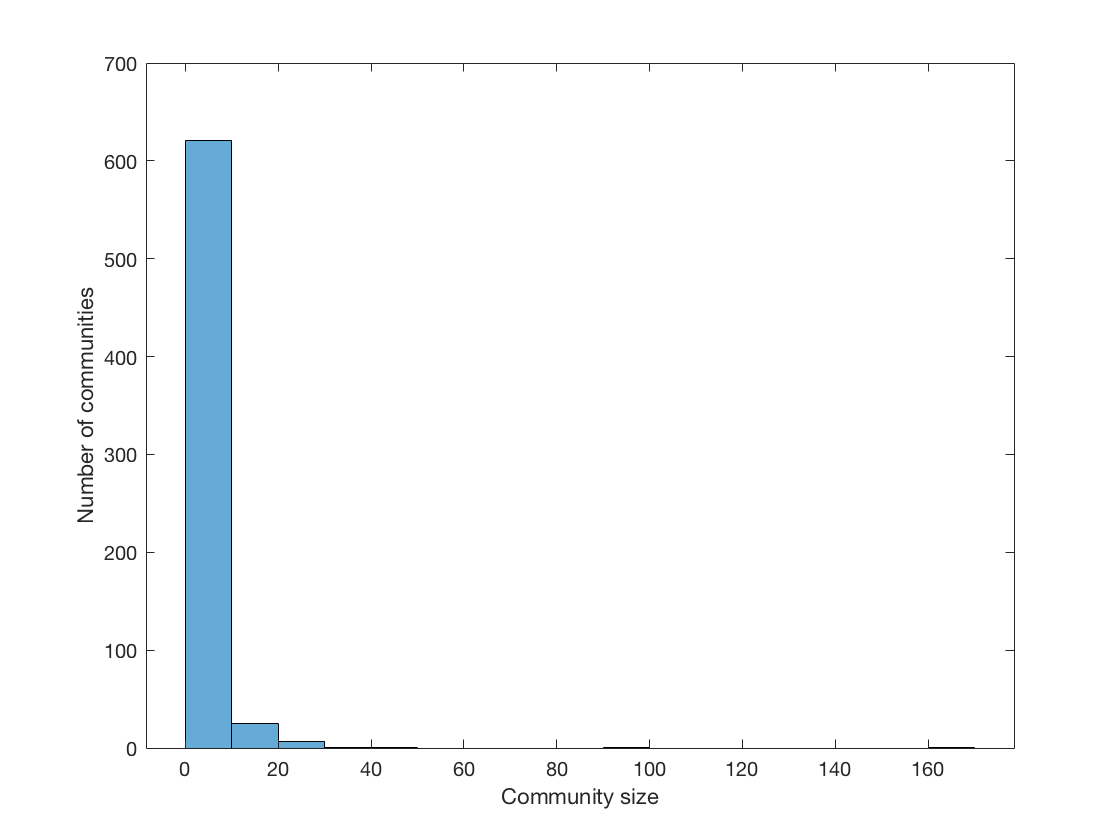}
	\caption{Size Distribution of Communities}
	\label{fig10}
\end{figure}

\begin{figure}[!t]
\centering
	\includegraphics[scale=0.25]{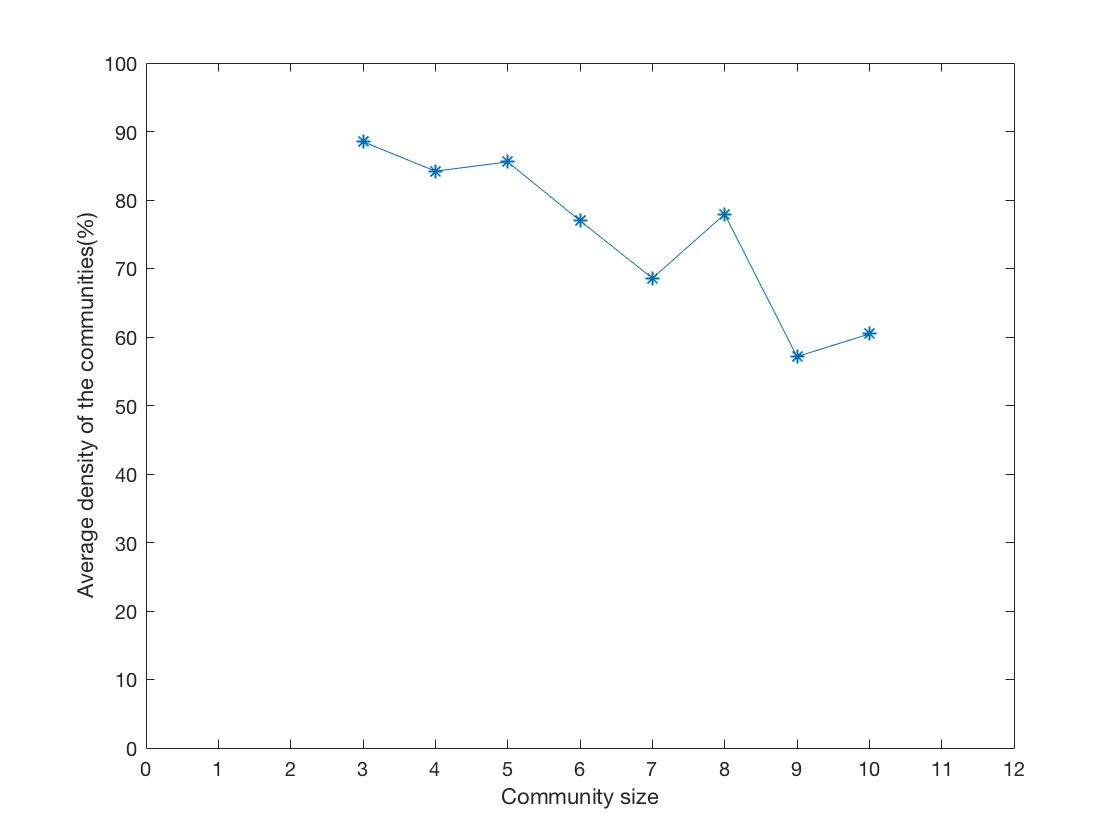}
	\caption{Average Network Density}
	\label{fig11}
\end{figure}

To further explore the social networks, all the candidates are integrated into communities with different size. Figure~\ref{fig10} shows the size distribution of all communities. Although the largest network size is 167, most of the social network sizes predicted using image data in this paper contain fewer than 10 persons. Therefore, the analysis of network density focuses on the network size of 3 to 10. The average network density of each size is shown in Figure~\ref{fig11}. Although the network density gradually decreases as the network size expands, the minimum network density is still 57.14$\%$.

To explore the features of image-generated community, a comparison between the communities predicted by the proposed technique and other social network communities built in \cite{Hashmi} is conducted. These communities include Twitter Friendship Network, Epinions Social Network, Wikipedia Vote Network and EU Email Communication Network. These four networks were constructed using textual data, such as user profiles, emails, and questionnaires.

Table~\ref{table5} shows the statistics of network properties calculated on our image-generated community set and four text-generated networks. Intuitively, the communities predicted using PIPA dataset have smaller sizes and higher densities than other four networks. Because small network size and high network density lead to cohesive communities, this indicates that the predicted communities are cohesive.

\begin{table}
\centering

\begin{tabular}{@{}cccccc@{}}
\toprule
Property& Twitter& Epinions& Wekipedia& Email& Image-generated \\ \midrule
Size& 500& 500& 500& 500& 3-10  \\
Edges& 3099& 13739& 11672& 2396 & 2-45 \\
Density& 6.18& 27.47& 23.34& 4.79& 88.51-57.14 \\ \bottomrule

\end{tabular}
\vspace{2mm}
\caption{Statistics for Comparison of Image-generated Network and Text-generated Network}
\label{table5}
\end{table}

\section{Conclusion and Future Work}
In this paper, we propose a deep learning based social network prediction system, CommuNety. The input to our deep neural network is an image of a target person, and the output is a target-centered predicted social network, which also presents the relationship strength of persons in the predicted social network. 

Due to lack of labeled image data and hardware limitation, data augmentation is used for the training of the proposed face recognition model. The training data is augmented by image rotation and all CNN features of images are exported from fc6 layer of the deep learning model. The features are fed to three fully connected layers to train the face recognition system. To predict and build social networks, face co-occurrence frequency technique is proposed to recognize people in the dataset who are directly or indirectly related to the target, and at the same time, use the face recognition model to classify each person’s identity. As the deep neural network is the core of the social network prediction system, hence its classification accuracy limits the performance of the system. We also propose a photo ranking algorithm to rank photos in the data set based on the TF-IDFs of persons in the same photos. Consequentially, relationship strength of identities in social network depends not only on the number of group photos, but also on the scores of these photos. This information is more valuable than simply constructing a social network. In addition, the social networks predicted using image data are smaller and more cohesive than other social networks.

In our future work, we aim to optimize and further improve our social network prediction system. We will consider using more complex deep learning model architectures and generate more training examples. Moreover, we also intend to explore other valuable information such as text and location of individuals from social networks and use them as additional features to improve the prediction of our proposed CommuNety system.

\section*{Acknowledgment}
This research is supported by Murdoch University, Australia. The authors would like to thank Dr Ammar Mahmood for useful discussion regarding face recognition and deep learning.

\ifCLASSOPTIONcaptionsoff
  \newpage
\fi



%\bibliographystyle{IEEEtran}
 
\bibliographystyle{IEEEtran}
\bibliography{mybibfile}

\begin{thebibliography}{10}
\providecommand{\url}[1]{#1}
\csname url@samestyle\endcsname
\providecommand{\newblock}{\relax}
\providecommand{\bibinfo}[2]{#2}
\providecommand{\BIBentrySTDinterwordspacing}{\spaceskip=0pt\relax}
\providecommand{\BIBentryALTinterwordstretchfactor}{4}
\providecommand{\BIBentryALTinterwordspacing}{\spaceskip=\fontdimen2\font plus
\BIBentryALTinterwordstretchfactor\fontdimen3\font minus
  \fontdimen4\font\relax}
\providecommand{\BIBforeignlanguage}[2]{{%
\expandafter\ifx\csname l@#1\endcsname\relax
\typeout{** WARNING: IEEEtran.bst: No hyphenation pattern has been}%
\typeout{** loaded for the language `#1'. Using the pattern for}%
\typeout{** the default language instead.}%
\else
\language=\csname l@#1\endcsname
\fi
#2}}
\providecommand{\BIBdecl}{\relax}
\BIBdecl

\bibitem{Ortiz}
E.~G. Ortiz and B.~C. Becker, ``Face recognition for web-scale datasets,''
  \emph{ELSEVIER Computer Vision and Image Understanding}, vol. 118, pp.
  153--170, 2014.

\bibitem{sun2016social}
L.~Sun, X.~Wang, Z.~Wang, H.~Zhao, and W.~Zhu, ``Social-aware video
  recommendation for online social groups,'' \emph{IEEE Transactions on
  Multimedia}, vol.~19, no.~3, pp. 609--618, 2016.

\bibitem{weng2009rolenet}
C.-Y. Weng, W.-T. Chu, and J.-L. Wu, ``Rolenet: Movie analysis from the
  perspective of social networks,'' \emph{IEEE Transactions on Multimedia},
  vol.~11, no.~2, pp. 256--271, 2009.

\bibitem{oro2017detecting}
E.~Oro, C.~Pizzuti, N.~Procopio, and M.~Ruffolo, ``Detecting topic
  authoritative social media users: a multilayer network approach,'' \emph{IEEE
  Transactions on Multimedia}, vol.~20, no.~5, pp. 1195--1208, 2017.

\bibitem{Pfeil}
U.~Pfeil, R.~Arjan, and P.~Zaphiris, ``Age differences in online social
  networking – a study of user profiles and the social capital divide among
  teenagers and older users in myspace,'' \emph{Computers in Human Behavior},
  pp. 643--654, 2009.

\bibitem{zhao2017social}
Z.~Zhao, Q.~Yang, H.~Lu, T.~Weninger, D.~Cai, X.~He, and Y.~Zhuang,
  ``Social-aware movie recommendation via multimodal network learning,''
  \emph{IEEE Transactions on Multimedia}, vol.~20, no.~2, pp. 430--440, 2017.

\bibitem{xu2018trust}
L.~Xu, T.~Bao, L.~Zhu, and Y.~Zhang, ``Trust-based privacy-preserving photo
  sharing in online social networks,'' \emph{IEEE Transactions on Multimedia},
  vol.~21, no.~3, pp. 591--602, 2018.

\bibitem{cheung2019detecting}
M.~Cheung and J.~She, ``Detecting social signals in user-shared images for
  connection discovery using deep learning,'' \emph{IEEE Transactions on
  Multimedia}, 2019.

\bibitem{garg2019hybrid}
S.~Garg, K.~Kaur, N.~Kumar, and J.~J. Rodrigues, ``Hybrid deep-learning-based
  anomaly detection scheme for suspicious flow detection in sdn: A social
  multimedia perspective,'' \emph{IEEE Transactions on Multimedia}, vol.~21,
  no.~3, pp. 566--578, 2019.

\bibitem{lu2016tag}
D.~Lu, X.~Liu, and X.~Qian, ``Tag-based image search by social re-ranking,''
  \emph{IEEE Transactions on Multimedia}, vol.~18, no.~8, pp. 1628--1639, 2016.

\bibitem{zhang2019personalized}
J.~Zhang, Y.~Yang, L.~Zhuo, Q.~Tian, and X.~Liang, ``Personalized
  recommendation of social images by constructing a user interest tree with
  deep features and tag trees,'' \emph{IEEE Transactions on Multimedia},
  vol.~21, no.~11, pp. 2762--2775, 2019.

\bibitem{Dmr}
DMR, ``Digital company statistics,''
  \url{https://expandedramblings.com/index.php/category/stats-2/digital-company-statistics/
  }, 2018.

\bibitem{Simonyan}
K.~Simonyan and A.~Zisserman, ``Very deep convolutional networks for
  large-scale image recognition,'' \emph{CoRR}, 2014.

\bibitem{Parkhi}
O.~M. Parkhi, A.~Vedaldi, and A.~Zisserman, ``Deep face recognition,''
  \emph{British Machine Vision Conference}, 2015.

\bibitem{Raza}
A.~S. Razavian, H.~Azizpour, J.~Sullivan, and S.~Carlsson, ``Cnn features
  off-the-shelf: an astounding baseline for recognition,'' \emph{CoRR}, 2014.

\bibitem{shah2017efficient}
S.~A. Shah, U.~Nadeem, M.~Bennamoun, F.~Sohel, and R.~Togneri, ``Efficient
  image set classification using linear regression based image
  reconstruction,'' in \emph{Proceedings of the IEEE conference on computer
  vision and pattern recognition workshops}, 2017, pp. 99--108.

\bibitem{hu2019disentangled}
W.~Hu and H.~Hu, ``Disentangled spectrum variations networks for nir-vis face
  recognition,'' \emph{IEEE Transactions on Multimedia}, 2019.

\bibitem{zhang2018dynamic}
Z.~Zhang, J.~Han, E.~Coutinho, and B.~Schuller, ``Dynamic difficulty awareness
  training for continuous emotion prediction,'' \emph{IEEE Transactions on
  Multimedia}, vol.~21, no.~5, pp. 1289--1301, 2018.

\bibitem{Hsu}
R.~Hsu, M.~Abdel-Mottaleb, and A.~K. Jain, ``Face detection in color images,''
  \emph{IEEE Transactions on Pattern Analysis and Machine Intelligence}, pp.
  696--706, 2002.

\bibitem{Zhang}
N.~Zhang, M.~Paluri, Y.~Taigman, R.~Fergus, and L.~Bourdev, ``Beyond frontal
  faces: Improving person recognition using multiple cues,'' \emph{CoRR}, 2015.

\bibitem{Chen}
Y.~Y. Chen, W.~H. Hsu, and H.~Y.~M. Liao, ``Discovering informative social
  subgraphs and predicting pairwise relationships from group photos,''
  \emph{Proceedings of the 20th ACM international conference on Multimedia},
  pp. 669--678, 2012.

\bibitem{Kim}
H.~N. Kim, A.~E. Saddik, and J.~G. Jung, ``Leveraging personal photos to
  inferring friendships in social network services,'' \emph{Expert Systems with
  Applications}, pp. 6955--6966, 2012.

\bibitem{Oh}
S.~J. Oh, R.~Benenson, M.~Fritz, and B.~Schiele, ``Person recognition in social
  media photos,'' \emph{CoRR}, 2017.

\bibitem{Dong}
Y.~Dong, Y.~Liu, and S.~Lian, ``Automatic age estimation based on deep learning
  algorithm,'' \emph{Neurocomputing}, pp. 4--10, 2016.

\bibitem{Geron}
A.~Geron, \emph{Hands-On Machine Learning with Scikit-Learn and
  TensorFlow}.\hskip 1em plus 0.5em minus 0.4em\relax O'Reilly Media, Inc.
  ©2017, 2017.

\bibitem{viola2004robust}
P.~Viola and M.~J. Jones, ``Robust real-time face detection,''
  \emph{International journal of computer vision}, vol.~57, no.~2, pp.
  137--154, 2004.

\bibitem{Ramos}
J.~Ramos, ``Using tf-idf to determine word relevance in document queries,''
  \url{http://www.cs.rutgers.edu/~mlittman/courses/ml03/iCML03/papers/ramos.pdf
  }, 2003.

\bibitem{brunelli1995person}
R.~Brunelli and D.~Falavigna, ``Person identification using multiple cues,''
  \emph{IEEE transactions on pattern analysis and machine intelligence},
  vol.~17, no.~10, pp. 955--966, 1995.

\bibitem{Candan}
S.~Candan, L.~Chen, T.~B. Pedersen, L.~Chang, and W.~Hua, \emph{Database
  Systems for Advanced Applications}.\hskip 1em plus 0.5em minus 0.4em\relax
  Springer International Publishing, 2017.

\bibitem{Hashmi}
A.~Hashmi, F.~Zaidi, A.~Sallaberry, and T.~Mehmood, ``Are all social networks
  structurally similar? a comparative study using network statistics and
  metrics,'' \emph{CoRR}, 2013.

\end{thebibliography}

\begin{IEEEbiography}[{\includegraphics[width=1in,height=1.25in,clip,keepaspectratio]{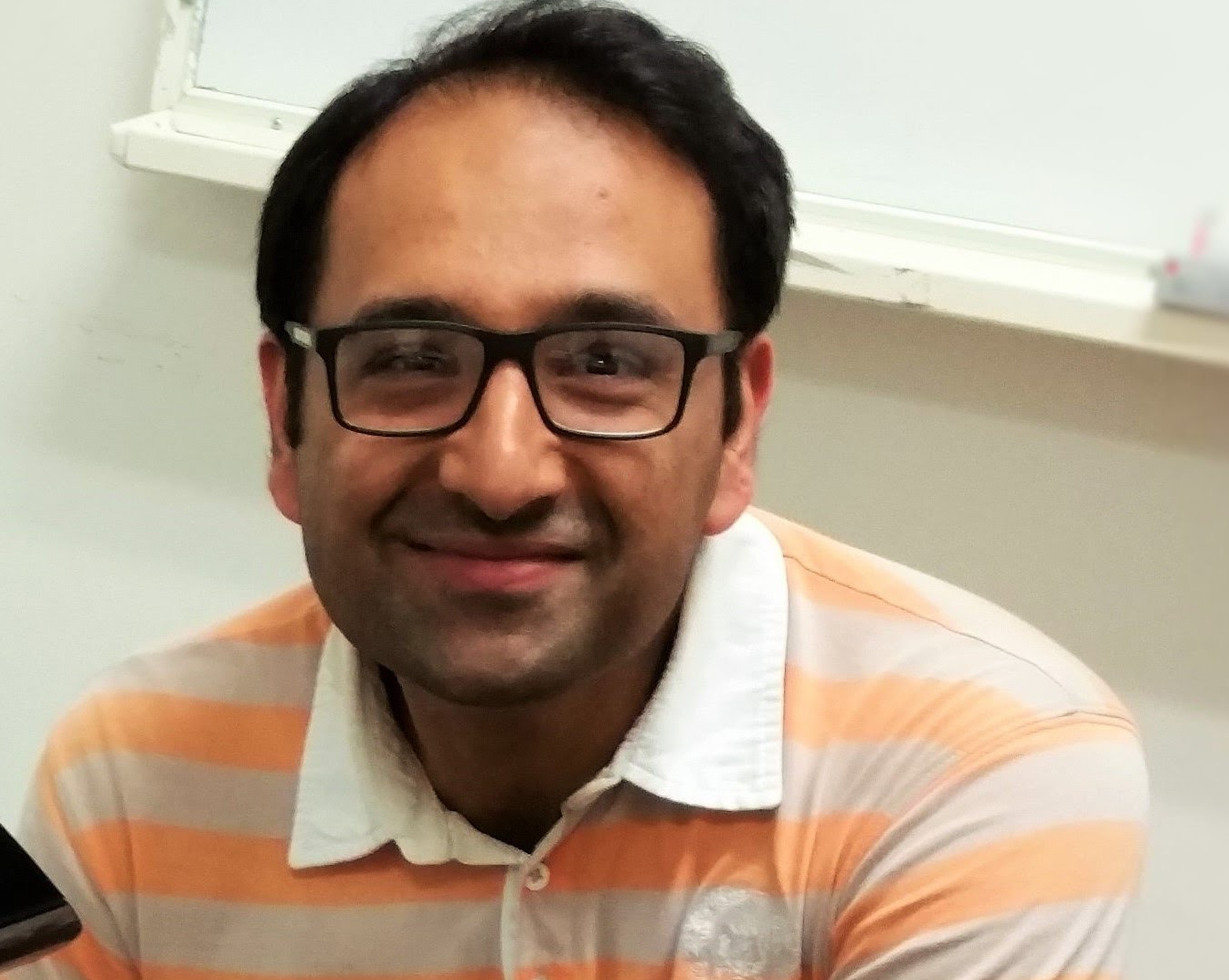}}]{Afaq Shah} received the Ph.D. degree
in computer vision and machine learning from The
University of Western Australia (UWA), Crawley,
WA, Australia.
He was a Lecturer ICT with Central Queensland
University, QLD, Australia. He is
currently a Lecturer with Murdoch University, Perth,
WA, Australia. He is also an Adjunct Lecturer with
the Department of Computer Science and Software
Engineering, UWA, Perth, WA, Australia. His current research interests include deep learning, computer vision, robotics, 3D object/face recognition and image processing. Dr. Shah was a recipient of the Start Something Prize for Research Impact through Enterprise for 3-D Facial Analysis Project funded by the Australian Research Council. He has co-authored a book, "A Guide  to Convolutional Neural Networks for Computer Vision".  
\end{IEEEbiography}

\begin{IEEEbiography}{Weifeng Deng} holds Masters by research degree from the University of Western Australia. His research interests include computer vision, image processing and deep learning.
\end{IEEEbiography}

\begin{IEEEbiography}[{\includegraphics[width=1in,height=1.25in,clip,keepaspectratio]{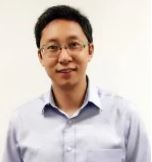}}]{Jianxin Li} is an A/Professor in the School of IT, Deakin University. His research interests include social computing, query processing and optimization, big data analytics, and intelligent computation in educational data. He has published 70 high quality research papers in top international conferences and journals, including PVLDB, IEEE ICDE, ACM WWW, IEEE ICDM, EDBT, ACM CIKM, IEEE TKDE, and ACM WWW. His professional service can be identified by different roles in academic committees, e.g., the technical program committee members in ACM SIGMOD, PVLDB, AAAI, PAKDD, IEEE ICDM, and ACM CIKM; the journal reviewer in IEEE TKDE, ACM TKDD, WWW Journal and VLDB Journal; the proceeding chairs in DASFAA 2018, ADMA 2016 and ADC 2015; and the program committee chair in the International Workshop on Social Computing 2017 and 2018; the tutorial chair in the 26th International Conference on WWW 2017; and the guest editors in international journals, such as Computational Intelligence, IET Intelligent Transport Systems, Complexity, Data Science and Engineering.
\end{IEEEbiography}

\begin{IEEEbiography}[{\includegraphics[width=1in,height=1.25in,clip,keepaspectratio]{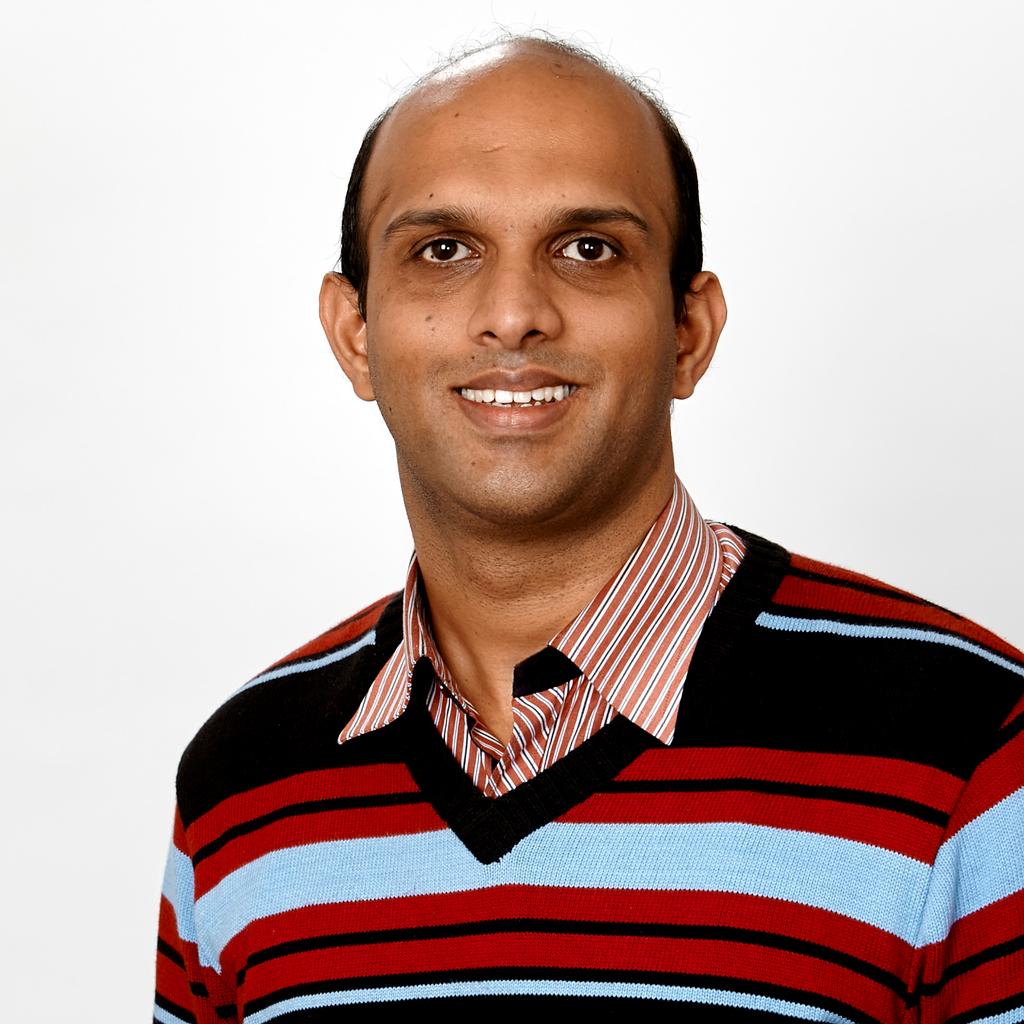}}]{Muhammad Aamir Cheema} is an ARC Future Fellow and an Associate Professor at Faculty of Information Technology, Monash University, Australia. He obtained his PhD from UNSW Australia in 2011. He is the recipient of 2012 Malcolm Chaikin Prize for Research Excellence in Engineering, 2013 Discovery Early Career Researcher Award, 2014 Dean’s Award for Excellence in Research by an Early Career Researcher, 2018 Future Fellowship, 2018 Monash Student Association Teaching Award and 2019 Young Tall Poppy Science Award. He has also won two CiSRA best research paper of the year awards, two invited papers in the special issue of IEEE TKDE on the best papers of ICDE, and three best paper awards at ICAPS 2020, WISE 2013 and ADC 2010, respectively. He is the Associate Editor of IEEE Transactions on Knowledge and Data Engineering. He served as PC co-chair for ADC 2015, ADC 2016, 8th ACM SIGSPATIAL Workshop ISA 2016 \& 2018, IWSC 2017, proceedings chair for DASFAA 2015 \& ICDE 2019, tutorial co-chair for APWeb 2017 \& MDM 2019 and publicity co-chair for ACM SIGSPATIAL 2017 \& 2018.

\end{IEEEbiography}

\begin{IEEEbiography}[{\includegraphics[width=1in,height=1.25in,clip,keepaspectratio]{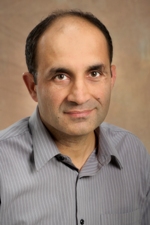}}]{Abdul Bais} received the M.Sc. degree in electrical engineering from University of Engineering and Technology, Peshawar, Pakistan, in 2003, and the Ph.D. degree in electrical engineering and information technology from Vienna University of Technology, Vienna, Austria, in 2007. 
From 2010 to 2013, he was a postdoctoral fellow with the Faculty of Engineering and Applied Science, University of Regina, Saskatchewan, Canada. Since 2015, he has been an Assistant Professor with the electronic systems engineering program at the Faculty of Engineering and Applied Science, University of Regina, SK, Canada. His research interests include signal processing, computer vision, deep learning, and big data analytics. He is a Licensed Professional Engineer in Saskatchewan, Canada.

\end{IEEEbiography}
\end{document}